\documentclass[3p,times,twocolumn]{elsarticle}

\usepackage{ecrc}
\usepackage{amsmath,amssymb}
\usepackage{tikz}
\usepackage{cancel}
\usepackage{tikzfeynman}

\volume{00}

\firstpage{1}

\journalname{Nuclear and Particle Physics Proceedings}

\runauth{}

\jid{nppp}

\jnltitlelogo{Nuclear and Particle Physics Proceedings}

\usepackage{amssymb}

\usepackage[figuresright]{rotating}

\begin{document}

\begin{frontmatter}



\dochead{}

\title{A flavor-safe composite explanation of $R_K$}


	\author{Adri\'an Carmona \corref{spk}}
	\author{Florian Goertz}
	\cortext[spk]{Speaker}

\address{CERN, Theoretical Physics Department,\\ CH-1211 Geneva 23, Switzerland}

\begin{abstract}
	In these proceedings we discuss a flavor-safe explanation of the  anomaly found in $R_K= {\cal B}(B \to K \mu^+ \mu^-)/{\cal B}(B \to K e^+ e^-)$ by LHCb, within the framework of composite Higgs models. 	We present a model featuring a non-negligible degree of compositeness for all three generations of right-handed leptons, which leads to a violation of lepton-flavor universality in neutral current interactions while other constraints from quark- and lepton-flavor physics are met. Moreoever, the particular embedding of the lepton sector considered in this setup provides a parametrically enhanded contribution to the Higgs mass that can weak considerably the need for ultra-light top partners. 
\end{abstract}


\end{frontmatter}


\section{Introduction}
\label{sec:intro}
Composite Higgs models provide an elegant explanation to the hierarchy problem by protecting the Higgs mass by its finite size \cite{Kaplan:1983fs, Georgi:1984af}. In addition, a sizable mass gap between the electroweak (EW) and the compositeness scale $\Lambda\approx 4\pi f_{\pi}$  can be achieved if one assumes the Higgs to be a pseudo Nambu-Goldstone boson (pNGB) of some global symmetry of the strong sector  \cite{Dimopoulos:1981xc, Contino:2003ve, Agashe:2004rs}. One typical assumption is that this global symmetry is only broken by the weak couplings of the elementary SM-like degrees of freedom, corresponding to the SM fermions -- with the possible exception of the right-handed (RH) top quark --   and gauge bosons, which generates a Higgs potential radiatively and triggers the electroweak symmetry breaking (EWSB). Within the paradigm of \emph{partial compositeness}, where one assumes linear mixings of the SM-like fermions with their composite counterparts,  the light mass eigenstates become mixtures of elementary and composite degrees of freedom,  tying together the dynamics behind the observed flavor pattern and EWSB. Since the Yukawa couplings are generated through such linear couplings after integrating out the corresponding composite counterparts, it is usually thought that only third generation quarks will exhibit a sizable degree of compositeness and will be relevant for EWSB. However, the fact that neutrinos may have Majorana masses, together with the observed non-hierarchical mixing pattern  in the PMNS matrix, can change this situation for the lepton sector, see e.g. \cite{delAguila:2010vg, Carmona:2014iwa}. In these proceedings we will discuss a very minimal implementation of leptons in composite Higgs models, where neutrino masses are generated via a type-III seesaw mechanism and the RH lepton sector is unified by embedding  the RH charged leptons and the RH neutrinos in a 
{\it single} representation of the global group $\mathcal{G}$  (for each generation)  \cite{Carmona:2015ena}. Linked to this unification, our setup predicts a violation of lepton-flavor  universality (LFU) in neutral current interactions, while LFU is basically respected
in charged currents, providing a natural and compelling explanation for the  $2.6\,\sigma$ deviation observed by LHCb \cite{Aaij:2014ora} in the very clean ratio \cite{Hiller:2003js, Guevara:2015pza, Bordone:2016gaq}
 
\begin{align}
R_K= \left. \frac{{\cal B}(B \to K \mu^+ \mu^-)}{{\cal B}(B \to K e^+ e^-)}\right|_{q^2 
\in [1,6]\,{\rm GeV}}^{\rm exp}=0.745^{+0.090}_{-0.074}\pm 0.036\,.
\end{align}
As we will see, this can be done in a completely flavor-save manner, due to the possibility of implementing a very economical flavor symmetry, which avoids the appearance of new sources of flavor-changing neutral currents (FCNC) to very good approximation. Since the lepton sector features a sizable degree of compositeness and the RH lepton unification requires the presence of non-minimal representations of $\mathcal{G}$, it will provide a parametrically enhanced correction to the Higgs mass, such that the need for ultra-light top partners is weakened considerably, linking the mass of the latter with the size of the neutrino masses.

\section{Setup}
\label{sec:setup}
Let us consider the so-called minimal composite Higgs model (MCHM), where the global symmetry of the strong sector $\mathcal{G}=SO(5)$ is broken by the strong dynamics to $\mathcal{H}=SO(4)$, delivering four Goldstone bosons that will be identified with the Higgs doublet.  
We consider the {\it minimal} custodial embedding of the SM lepton sector including three RH fermion triplets with zero hypercharge,  $\Sigma_{\ell R}$, with $\ell=e,\mu,\tau$. If these new degrees of freedom have Majorana masses of order $\mathcal{O}(M_{\rm GUT})$,  the observed tiny neutrino masses can be explained with $\mathcal{O}(1)$ Yukawa couplings via the (type-III) \emph{seesaw} mechanism. In the framework of the MCHM, or its five dimensional (5D) holographic dual \cite{Maldacena:1997re,Gubser:1998bc,Witten:1998qj,ArkaniHamed:2000ds}, this is realized  by embedding every generation of RH leptons in a symmetric representation ($\mathbf{14}$) of $SO(5)$, whereas every left-handed (LH) doublet is embedded in a fundamental representation ($\mathbf{5}$) of $\mathcal{G}$. In terms of the different 5D bulk fields transforming under $SO(5)\times U(1)_X$, such embedding of the lepton sector reads $\zeta_1^{\ell}\sim \mathbf{5}_{\mathbf{-1}}$ and $\zeta_2^{\ell}\sim\mathbf{14}_{-1}$, for $\ell=e,\mu,\tau$,\,\footnote{For simplicity, we will be rather schematic in the description of the 5D setup. We thus refer the reader to Ref.\,\cite{Carmona:2015ena} for further details.}
\begin{align}
\zeta_{1}^{\ell}&=\ell^{\prime}_{1}[-,+]\oplus \left(\begin{array}{r}\nu_{1}^{\ell}[+,+]~  ~\tilde{\ell}_1[-,+]\\ \ell_{1}[+,+]~ \tilde{Y}_1^{\ell}[-,+]\end{array}\right),\nonumber\\
\zeta_{2}^{\ell}&=\ell^{\prime}_{2}[-,-]\oplus\left(\begin{array}{r} \nu_{2}^{\ell}[+,-]~~  \tilde{\ell}_2[+,-]\\ \ell_{2}[+,-]~ \tilde{Y}_2^{\ell}[+,-]\end{array}\right) \\
						   &\oplus\left(\begin{array}{r} \hat{\lambda}^{\ell}_2[-,-]~~\nu_{2}^{\ell\prime\prime}[+,-]~  ~~\ell_2^{\prime\prime\prime}[+,-]\\ 
	\hat{\nu}_2^{\ell}[-,-]~~~\ell_{2}^{\prime\prime}[+,-]~Y_2^{\ell\prime\prime\prime}[+,-]\\ \hat{\ell}_2[-,-]~Y_2^{\ell\prime\prime}[+,-]~\Theta_2^{\ell\prime\prime\prime}[+,-] \end{array}\right),\nonumber\qquad
\end{align}
where  we have explicitly shown the decomposition under $SU(2)_L\times SU(2)_R\cong SO(4)=\mathcal{H}$ (with the bidoublet being represented by a $2\times 2$ matrix on which the $SU(2)_L$ rotation acts vertically and the $SU(2)_R$ one horizontally)  and the signs in square brackets denote the boundary conditions at the UV and IR branes. A Dirichlet boundary condition for the RH/LH chirality is denoted by $[+/-]$, with LH/RH zero modes being present for fields with $[+,+]/[-,-]$ boundary conditions.  Finally, since the lepton sector will produce an additional non-negligible contribution to the Higgs potential, we can consider for the quark sector the previously disregarded minimal model consisting of  a fully composite $t_R$ and a  LH doublet $q_L^3$ embedded in a $\mathbf{5}$ of $\mathcal{G}$. More specifically, we consider $\xi_1^{i}\sim \mathbf{5}_{2/3},\, \xi_2^i\sim \mathbf{1}_{2/3},\, \xi_3^{i}\sim \mathbf{5}_{-1/3},\,  \xi_4^i\sim \mathbf{1}_{-1/3}$,  $i=1,2,3$, or
\begin{align}
	&\xi_{1}^i=\left(\begin{array}{r}\tilde{\Lambda}^i[-,+]~  u^i_1[+,+]\\ \tilde{u}^i[-,+]~ d_1^i[+,+]\end{array}\right)\oplus u_1^{i\prime}[-,+],\nonumber\\ &\xi_{2}^i[-,-],\\
		&	\xi_{3}^i=\left(\begin{array}{r}u^i_3[-,+]~\tilde{d}^i[-,+]  \\ d_3^i[-,+]~\tilde{\Xi}^i[-,+]\end{array}\right)\oplus d_3^{i\prime}[-,+],\nonumber\\ &\xi_{4}^i[-,-].\nonumber
\end{align}

This minimal realization of composite leptons naturally allows for a very strong  flavor protection, requiring any lepton flavor violating (LFV) process to be mediated by extremely suppressed neutrino-mass insertions and leading in particular to the absence of dangerous FCNCs in the lepton sector to excellent approximation. To this end, we promote the accidental $SU(3)_1\times SU(3)_2$ flavor symmetry of the lepton sector in the decompactified or conformal limit (arising from the arbitrary rotation of $\xi_{1}$ and $\xi_2$ in the family space) to a 5D gauge group only broken at the UV brane (i.e., by the elementary sector) and the vacuum expectation value (vev) of some non-dynamical field $\mathcal{Y}$  \cite{Fitzpatrick:2007sa, Perez:2008ee}.  The bulk fields in the lepton sector will thus transform as $\zeta_1\sim (\mathbf{3},\mathbf{1})$ and $\zeta_2\sim (\mathbf{1},\mathbf{3})$, whereas  $\mathcal{Y}\sim (\mathbf{3},\mathbf{\bar{3}})$. Therefore, the corresponding bulk masses will be given by\,\footnote{The $\ldots$ stand for subleading contributions $\mathcal{Y}\mathcal{Y}^{\dagger}\mathcal{Y}\mathcal{Y}^{\dagger}$, $\mathcal{Y}^{\dagger}\mathcal{Y}\mathcal{Y}^{\dagger}\mathcal{Y}, \ldots,$ which do not add additional flavor structure since they can all be made diagonal by (\ref{eq:rot}).}
\begin{align}
	c_1=\eta_1\mathbf{1}+\rho_1\mathcal{Y}\mathcal{Y}^{\dagger}+\ldots,\qquad c_2=\eta_2\mathbf{1}+\rho_2\mathcal{Y}^{\dagger}\mathcal{Y}+\ldots,
	\label{eq:bm}
\end{align}
whereas the IR brane masses will read
\begin{align}
	\left.a^4\left[ \omega_S\left(\bar{\zeta}_{1L}^{(\mathbf{1},\mathbf{1})}\mathcal{Y}\zeta_{2R}^{(\mathbf{1},\mathbf{1})}\right)+\omega_B(\bar{\zeta}_{1L}^{(\mathbf{2},\mathbf{2})}\mathcal{Y}\zeta_{2R}^{(\mathbf{2},\mathbf{2})})\right]\right|_{R^{\prime}}+\mathrm{h.c.}, 
	\label{eq:brm}
\end{align}
with $\eta_{1,2},\rho_{1,2}\in\mathbb{R}$, $\omega_{S,B}\in\mathbb{C},$ $a(z)=R/z$ the warp factor, $z\in[R, R^{\prime}]$ the coordinate of the extra dimension and the superscripts $(\mathbf{1},\mathbf{1})$ and $(\mathbf{2},\mathbf{2})$ denoting the singlet and the bidoublet components of the corresponding multiplets. Since, as mentioned, the elementary sector represented by the UV brane does not respect in general this symmetry, one can have    general Majorana masses
\begin{align}
\mathcal{L}_{\rm UV}\supset-\frac{1}{2}\left.M_{\Sigma}^{\ell\ell^{\prime}}\mathrm{Tr}\left(\bar{\Sigma}_{\ell R}^c \Sigma_{\ell^{\prime} R}\right)\right|_{z=R}+\mathrm{h.c.},
	\label{eq:uvm}
\end{align}
where
\begin{eqnarray}
\Sigma_{\ell}=\begin{pmatrix}\hat{\nu}_2^\ell/\sqrt{2}&\hat{\lambda}_2^\ell\\ \ell_2& -\hat{\nu}_2^\ell/\sqrt{2}\end{pmatrix}, \qquad \ell=e,\mu,\tau\,.
\end{eqnarray} 
Note that the fact of having just two $SO(5)$ lepton multiplets and thus being able to use only one $SU(3)_1\times SU(3)_2$ spurion, $\mathcal{Y}$, allows us to diagonalize at the same time  (\ref{eq:bm}) and (\ref{eq:brm}) via the rotation
\begin{align}
	\zeta_1\to \mathcal{U}_1\zeta_1,\qquad  \zeta_2\to \mathcal{U}_2\zeta_2,
	\label{eq:rot}
\end{align}
where $\mathcal{U}_1^{\dagger}\mathcal{Y}~\mathcal{U}_2=\mathrm{diag}(y_{ee},y_{\mu\mu},y_{\tau\tau})\equiv y_{\ell\ell}$. In this particular basis, the whole Lagrangian will be flavor diagonal with the exception of the Majorana mass in (\ref{eq:uvm}), which becomes $\mathcal{U}_2^{T}M_{\Sigma} ~\mathcal{U}_2$.
Therefore, any  potentially induced FCNC will be suppressed by large Majorana masses. Regarding the quark sector, we consider the more general case of arbitrary sources of flavor breaking, see \cite{Carmona:2015ena} for more details.

\section{EWSB, lepton non-universality and $R_K$}
\label{sec:rk}
In order to make the discussion simpler, it will be useful in the following to use the language of the dual four dimensional (4D) strongly coupled theory. Very schematically, we consider an elementary sector, consisting of the would-be SM with the exception of the Higgs sector and the addition of the corresponding RH neutrinos $\Sigma_{\ell}$ needed for the see-saw mechanism, and a composite sector mixing linearly with the elementary one. Focusing on the lepton sector, this mixing will be given by the following Lagrangian 
	\begin{align}
		\mathcal{L}_{\rm mix}^{\rm lep}=\sum_{\ell}\frac{\lambda_{L}^{\ell}}{\Lambda^{\gamma_L^{\ell}}}\bar{l}_{\ell L}\mathcal{O}_{\ell L}+\sum_{\ell}\frac{\lambda_R^{\ell}}{\Lambda^{\gamma_R^{\ell}}}\bar{\Psi}_{\ell R}\mathcal{O}_{\ell R}+\mathrm{h.c.},
	\end{align}
where  $\Lambda=\mathcal{O}(M_{\mathrm{Pl}})$ is the UV cut-off scale, $\gamma_{L,R}^{\ell}=[\mathcal{O}_{\ell L,R}]-5/2$ are the different anomalous dimensions, $\lambda_{L,R}^{\ell}$ are order one dimensionless parameters and all RH leptons have been embedded in $\Psi_{\ell R}\sim \mathbf{14}$. Since we expect $\gamma_{R}^{\ell}<0$ due to the size of the neutrino masses (since otherwise the elementary Majorana mass  $M_{\Sigma}\sim \Lambda$ would generate too small neutrino masses $\sim v^2/\Lambda$), $\Psi_{\ell R}$ will be rather composite and a large contribution to the $\Psi_{\ell R}$ kinetic term will be generated at the scale $\mu=\mathcal{O}(\rm{TeV})$ where the conformal sector becomes strongly coupled,
\begin{align}
	\frac{\lambda_R^{\ell 2}}{\Lambda^{2\gamma_R^{\ell}}}\int \mathrm{d}^4p~ \mathrm{d}^4q~\bar{\Psi}_{\ell R} (-p) \langle\mathcal{O}_{\ell R}(p) \bar{\mathcal{O}}_{\ell^{\prime}R}(-q) \rangle   \Psi_{\ell^{\prime} R}(q)\nonumber\\
	\sim \delta_{\ell \ell^{\prime}}\lambda_R^{\ell 2}\left(\frac{\mu}{\Lambda}\right)^{2\gamma_R^{\ell}}\int \mathrm{d}^4x~\bar{\Psi}_{\ell R}(x)i\cancel{\partial}\Psi_{{\ell}R}(x).\end{align}
This leads, after canonical normalization, to the following expressions for the physical masses, 
\begin{align}
	\mathcal{M}_{e}\sim \delta_{\ell \ell^{\prime}} v\epsilon_{\ell L}\quad \mathcal{M}_{\nu}\sim v^2 \epsilon_{\ell L}\epsilon_{\ell R}\left(M_{\Sigma}\right)^{-1}_{\ell \ell^{\prime}} \epsilon_{\ell^{\prime} L}\epsilon_{\ell^{\prime} R}
\end{align}
where $v$ is the Higgs vev and we have defined $\epsilon_{\ell L,R}=\lambda_{L,R}^{\ell}(\mu/\Lambda)^{\gamma_{L,R}^{\ell}}$. It is then clear  that in order to have simultaneously hierarchical charged lepton masses and a non-hierarchical neutrino mass matrix one \emph{needs} 
\begin{align}
	\epsilon_{eL}\ll \epsilon_{\mu L}\ll \epsilon_{\tau L}\ll 1\quad \mathrm{and} \quad \epsilon_{\ell L} \epsilon_{\ell R}\sim \mathrm{constant}
\end{align}
and thus 
\begin{align}
	0 \ll \epsilon_{\tau R}\ll \epsilon_{\mu R}\ll \epsilon_{e R}.
	\label{eq:comp}
\end{align}

\begin{figure}[!h]
	\begin{center}
			\includegraphics[width=0.48\textwidth]{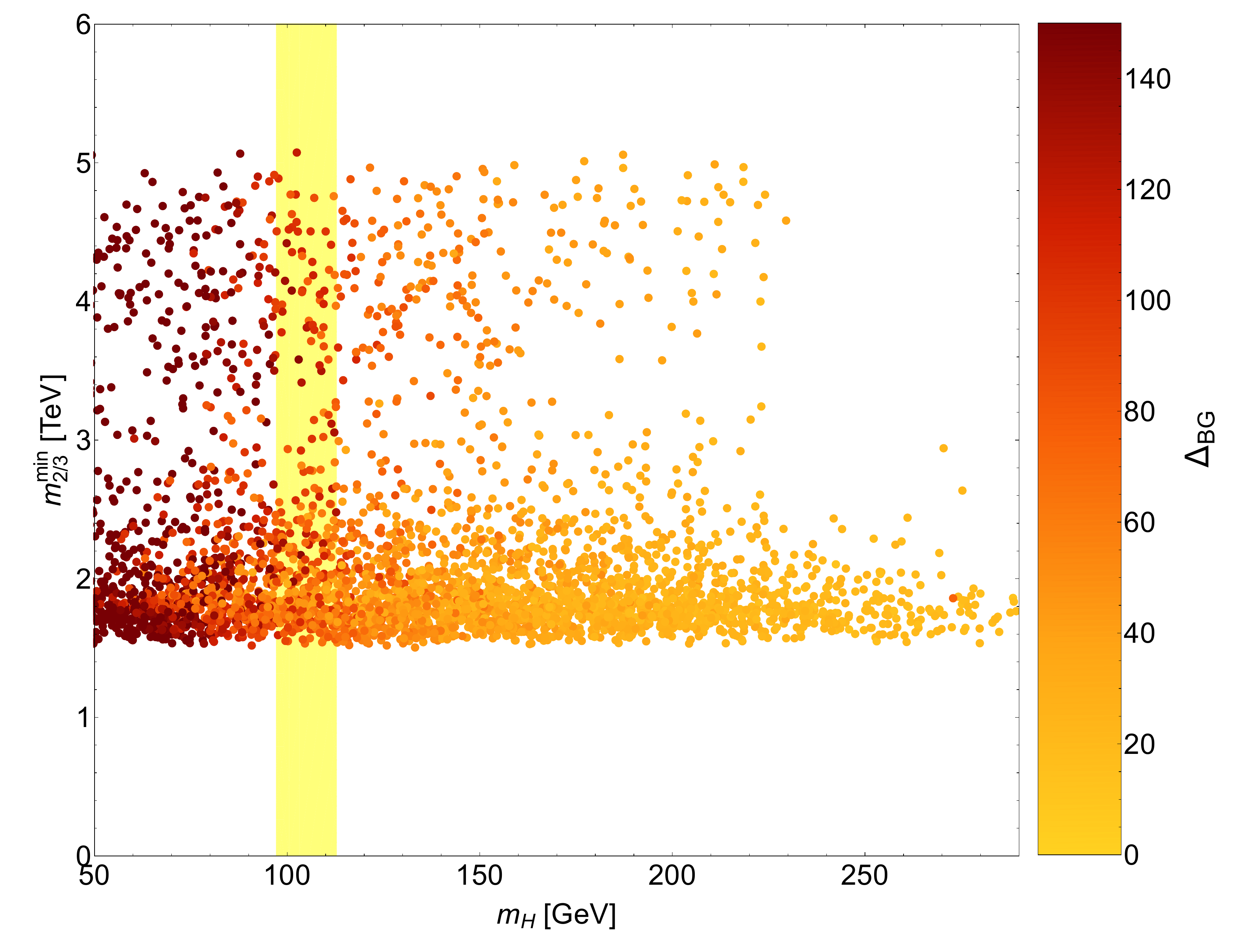}
		   \caption{Mass of the lightest top partner versus the Higgs mass as a function of the tuning $\Delta_{\rm BG}$, with lighter points corresponding to smaller values of $\Delta_{\rm BG}$, for $f_\pi=1\,$TeV. The yellow band corresponds to $m_H(f_\pi)=105\,\mathrm{GeV}~(1\pm7.5\%)$.}
			\label{fig:first}
		\end{center}
\end{figure}

This has several interesting consequences. First of all, since the degree of compositeness of RH leptons is non-negligable and their contribution to the Higgs quartic scale with $ \epsilon_{\ell R}^2$, instead of the usual $ \epsilon_{\ell R}^4$ for smaller representations of $\mathcal{O}_{\ell R}$, leptons give a sizable contribution to the Higgs potential.  As already mentioned, this is interesting since it allows to replace the role of $t_R$ in EWSB, making possible for $\ell_R$ to cancel the $q_L^3$ contribution with a moderate value of $\epsilon_{\ell R}$ such as to generate a viable potential.  Along the same lines, the lepton contribution can provide a negative correction to $m_H$, which in turn allows bigger masses for the top partners (which otherwise would drive $m_H$ too large \cite{Carmona:2015ena}). In order to get a more quantitative idea of the impact of the lepton sector on the Higgs potential and the Higgs mass, we show in Figure~\ref{fig:first} the mass of the lightest top partner versus the Higgs mass evaluated at the composite scale $\mathcal{O}(f_\pi)$,  with $f_\pi=1$\,TeV and the yellow band corresponding to the high-scale value of the actual Higgs mass $m_H(f_\pi)=105\,\textrm{GeV}~ (1\pm7.5\%)$, after accounting for the uncertainties of the running in a conservative way. We also display the Barbieri-Giudice (BG) measure of the tuning   $\Delta_{\rm BG}$, through the color of each point in the $m_{H}-m_{2/3}^{\rm min}$ plane. We can see from the figure that top-partner masses up to $5\,$TeV are allowed with a more than reasonable amount of tuning.

\begin{figure}[!h]
	\begin{center}
\begin{tikzpicture}[line width=1.5 pt, scale=.67]
	\draw[fermion] (-3,1.) -- (-2.25, 0.5);
	\draw[double] (-2.25,0.5) -- (-1.5, 0);
			\begin{scope}[shift={(-12:-2.25)}]
				\draw (125:.15) -- (-55:.15);
				\draw (55:.15) -- (-125:.15);			
			\end{scope}
	\draw[fermion] (-2.25,-0.5) -- (-3,-1.);
	\draw[double] (-1.5,0) -- (-2.25,-0.5);
			\begin{scope}[shift={(12:-2.25)}]
				\draw (125:.15) -- (-55:.15);
				\draw (55:.15) -- (-125:.15);			
			\end{scope}
		\draw [vector, double] (-1.5,0) -- (1.5,0);
		\draw[double] (1.5, 0) -- (2.25,0.5);
		\draw[fermion] (2.25, 0.5) -- (3,1);
			\begin{scope}[shift={(12:2.25)}]
				\draw (125:.15) -- (-55:.15);
				\draw (55:.15) -- (-125:.15);			
			\end{scope}
	\draw[fermion] (3,-1.) -- (2.25,-0.5);
	\draw[double] (2.25,-0.5) -- (1.5,0);
			\begin{scope}[shift={(-12:2.25)}]
				\draw (125:.15) -- (-55:.15);
				\draw (55:.15) -- (-125:.15);			
			\end{scope}
		\draw[fill=gray] (-1.5,0) circle (.2cm);
		\draw[fill=gray] (1.5,0) circle (.2cm);
		\node at (-3.5, 1) {$\psi_1$};
		\node at (-3.5, -1) {$\bar{\psi}_2$};
		\node at (3.5, 1) {$\chi_1$};
		\node at (3.5, -1) {$\bar{\chi}_2$};
	\end{tikzpicture}
	\caption{Relevant diagrams for the generation of four-fermion operators.}
	\label{fig:dg}
	\end{center}
\end{figure}
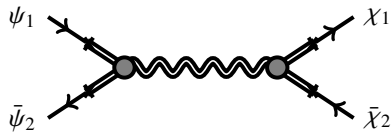

Secondly, since different RH leptons exhibit a \emph{different} degree of compositeness, see eq. (\ref{eq:comp}), diagrams with a tree-level exchange of neutral heavy vector resonances like the ones schematically depicted in Figure\,\ref{fig:dg} will lead to a violation of LFU.\,\footnote{See Refs.~\cite{Gripaios:2014tna, Niehoff:2015bfa, Buttazzo:2016kid} for different examples in the context of CHMs and Refs.~\cite{Crivellin:2015mga, Crivellin:2015lwa, Sierra:2015fma, Celis:2015ara, Falkowski:2015zwa, Celis:2015eqs, Boucenna:2016qad} for other $Z^{\prime}$ models.} In particular, they will lead to four-fermions operators 
\begin{align}
	c\mathcal{O}\sim c\, (\bar{\psi}_2\gamma_{\mu} \psi_1)(\bar{\chi}_2\gamma^{\mu} \chi_1), \quad c\sim \epsilon_{\psi_1}\epsilon_{\psi_2}\epsilon_{\chi_1}\epsilon_{\chi_2}/f^2_{\pi},
\end{align}
that, besides flavor, will be relevant also for electroweak precision data (EWPD).

\begin{figure}[!h]
	\begin{center}
			\includegraphics[width=0.42\textwidth]{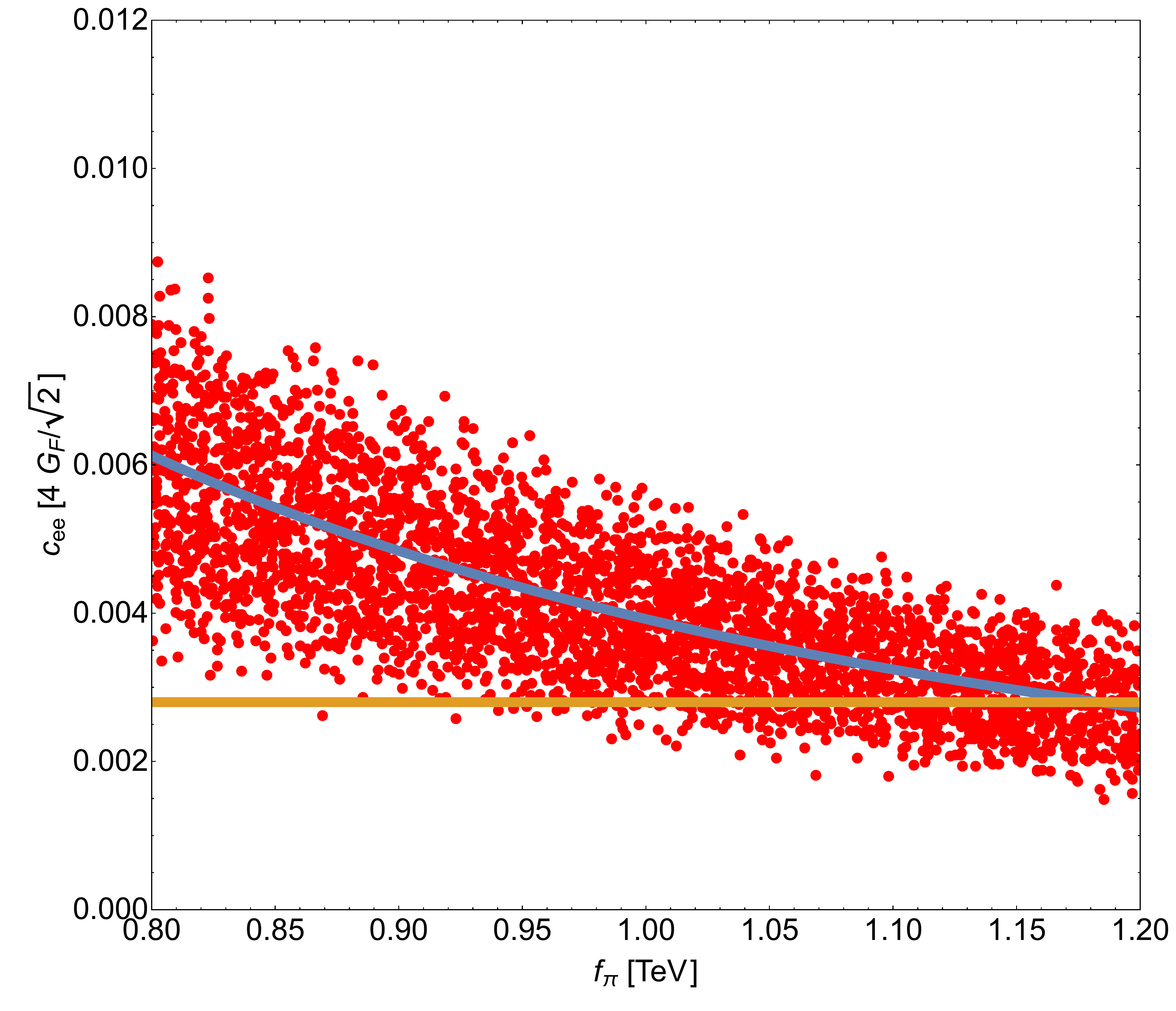}
			\caption{Value of the $c_{ee}$ Wilson coefficient  as a function of $f_\pi$. The blue curve shows the best fit to the data while the yellow line corresponds to the upper bound at 95\% C.L..}
			\label{fig:4fer}
		\end{center}
\end{figure}

According to eq. (\ref{eq:comp}), the most important of these operators regarding EWPD will be $\mathcal{O}_{ee}=(e_R\gamma_{\mu} e_R)(e_R\gamma^{\mu} e_R)/2$, whose Wilson coefficient $c_{ee}$  is constrained to be  $c_{ee}\in 4 G_F/\sqrt{2}\cdot [-1.8,+2.8] \cdot 10^{-3}$ at 95\% C.L.  \cite{Raidal:2008jk}.
We present in Figure~\ref{fig:4fer} the value of $c_{ee}$  as a function of $f_\pi$, where the blue curve corresponds to the best fit to the data. We also show the 95\% C.L. upper bound on $c_{ee}$ by a yellow line. One can see from this plot that values of $f_\pi\gtrsim 1\,$TeV give already a reasonable agreement with the data, while for $f_\pi\gtrsim 1.2\,$TeV the EWPD impose no significant constraint. Therefore, in order to provide a conservative assessment of the flavor predictions of the model, we consider $f_\pi=1.2\,$TeV henceforth.

Concerning flavor, the most relevant operators will be 
\begin{align}
	\mathcal{O}_{qe}^{32\ell \ell}&=\left(\bar{q}_{L}^2 \gamma_{\mu} q_L^3\right)\left(\bar{\ell}_R \gamma^{\mu} \ell_R\right), \\ \mathcal{O}^{B_s}_1&=\left(\bar{q}_L^2\gamma_{\mu} q_L^3\right)\left(\bar{q}_L^2\gamma^{\mu} q_L^3\right),
\end{align}
where the first of them  will provide the leading contribution to $R_K$ and we expect the latter to appear unavoidably if we generate the first one. Instead of performing a complete flavor analysis of the quark sector, we prefer to focus on the possible correlations between $B_s-\bar{B}_s$ mixing and $R_K$. 
On the other hand, note already that a large class of potentially dangerous constraints, coming from limits on LFU violation in charged current interactions,
mediating e.g. $K$, $\pi$, and $\mu$ decays \cite{Antonelli:2010yf,Greljo:2015mma}, is fulfilled in this model by construction. In fact, the LH charged current $\bar \ell_L \gamma_\mu \nu^\ell_L$ is mostly elementary and the light neutrino mass eigenstates contain only a negligible amount of RH fields.
Thus, charged currents respect LFU to excellent approximation.

We evaluate $R_K$ by computing the Wilson coefficients of the   
\begin{align}
	\mathcal{O}_{9(10)}^{\ell}&=\left[\bar{s}\gamma_{\alpha}P_Lb\right]\left[\bar{\ell}\gamma^{\alpha}(\gamma_5)\ell\right],\\
	\mathcal{O}_{9(10)}^{\prime \ell}&=\mathcal{O}_{9(10)}^{\ell}[P_L\to P_R]
\end{align}
operators from the usual $|\Delta B|=|\Delta S|=1$ Hamiltonian \cite{Ghosh:2014awa}. Note that, even though we are also generating contributions to $\mathcal{O}_{10}^{\ell}$ and $\mathcal{O}_{10}^{\prime \ell}$ that could in principle lead to large deviations with respect to the SM predictions in $B_s\to \ell^{+}\ell^{-}$ decays \cite{Bobeth:2013uxa}, 
 we  expect the largest effect to arise in the poorly measured $B_s\to e^{+}e^-$ decay, rather than in  $B_s\to \mu^{+}\mu^{-}$ \cite{CMS:2014xfa}. 

\begin{figure}[!t]
	\begin{center}
			\includegraphics[width=0.42\textwidth]{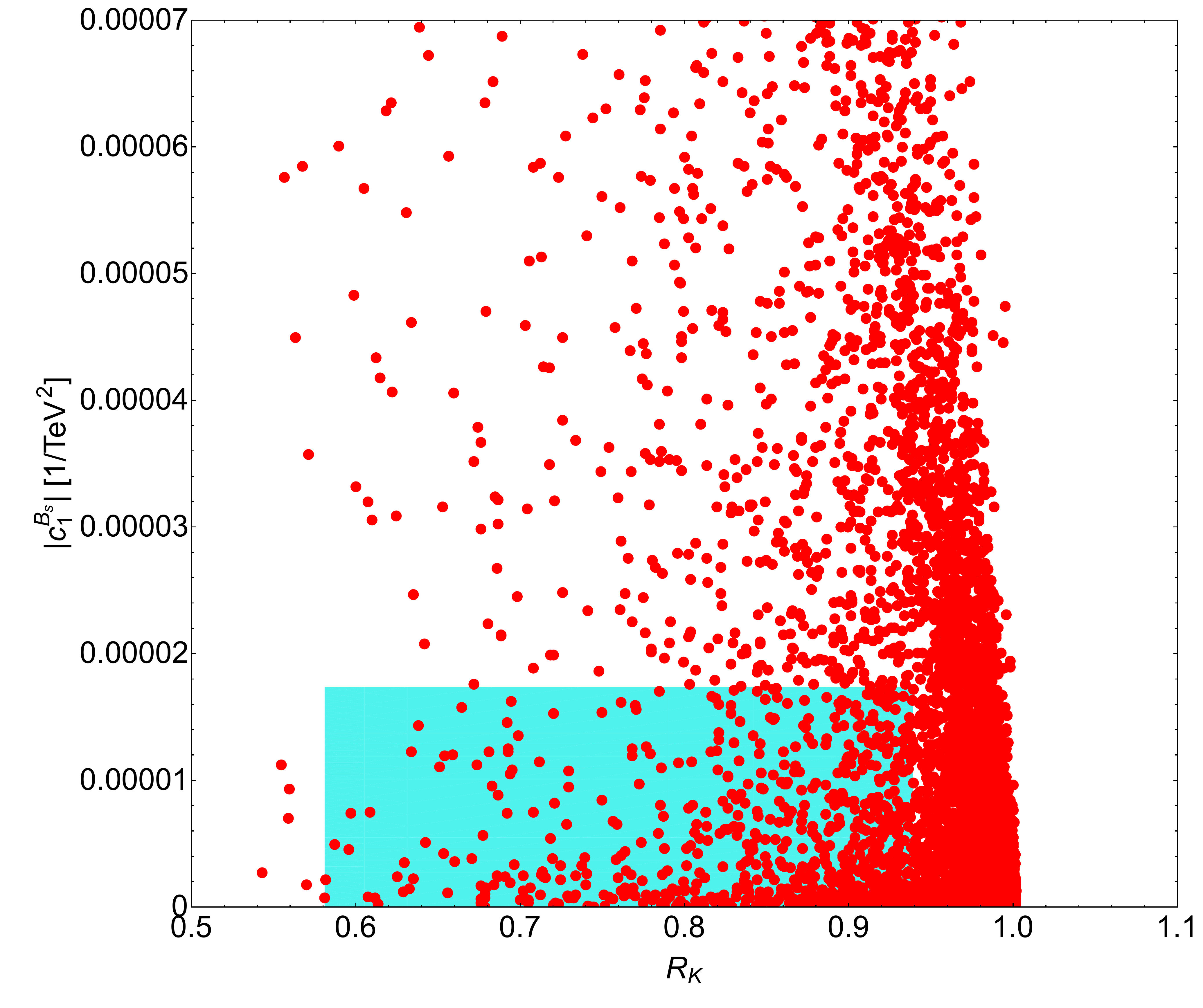}
			\caption{Value of $|c_{1}^{B_s}(m_{\rho})|$ versus $R_K$ for points reproducing the Higgs mass and within 2$\sigma$ from $B_s\to \mu^{+}\mu^-$, for $f_{\pi}=1.2\,$TeV. The blue box marks the allowed values of $R_K$ and $|c_{1}^{B_s}|$ at $95\%$ C.L..}
			\label{fig:RK}
		\end{center}
\end{figure}

We show in Figure~\ref{fig:RK} the values of $|c_{1}^{B_s}(m_{\rho})|$ versus $R_K$ for the points of the scan with the correct Higgs mass and a 2$\sigma$ agreement with the measured value of $B_s\to \mu^{+}\mu^{-}$ for  $f_{\pi}=1.2\,$TeV. The blue box represents the allowed values in the $R_K-|c_1^{B_s}|$ plane at 95\% C.L., taking into account  $|c_1^{B_s}|\le (240\, \mathrm{TeV})^{-2}$~\cite{Bevan:2014cya}.
It is clear from the plot that, even in the conservative case of $f_{\pi}=1.2\,$TeV, which guarantees the agreement with EWPD,  we can  explain the observed value of $R_K$ while not violating the bounds from  $B_s - \bar B_s$ mixing or $B_s\to \mu^{+}\mu^{-}$ for a sizable region of the parameter space. 

\section{Conclusions}
\label{sec:conc}
Flavor physics provides a superb tool for probing physics beyond the SM. It is therefore far from being a surprise that so much excitement has been raised by the $B-$physics anomalies observed both in the charged and the neutral currents. From all of them, $R_K$ stands out particularly since it provides a very clean probe of lepton flavor non-universality, only produced at the loop level in the SM. While there are plenty of models on the market explaining the latter, there are only a few of them which could be motivated from an UV perspective. In these proceedings, we have discussed a model where the observed deviation in  $R_K$ can be explained naturally within the context of composite Higgs setups, offering a beautiful link between naturalness and the violation of LFU. What is more, this can be achieved in a completely flavor-safe way,  avoiding potentially dangerous FCNCs in the lepton sector, while alleviating at the same time the necessity of ultra-light top partners. Therefore, if confirmed, the final observation of violation of LFU could provide an unexpected first probe of the dynamics solving the hierarchy problem. 

\begin{paragraph}{Acknowledgments}
The research of A.C. has been supported by a Marie Sk\l{}odowska-Curie Individual Fellowship of the European Community's Horizon 2020 Framework Programme for Research and Innovation under contract number 659239 (NP4theLHC14). The research of F.G. is supported by a Marie Curie Intra European Fellowship within the EU FP7 (grant no. PIEF-GA-2013-628224). 
\end{paragraph}



\bibliographystyle{elsarticle-num}
\bibliography{myrefs}







\end{document}